\documentstyle[psfig]{caosp}
\begin{document}
%
%
\pubyear{1998}
\volume{23}
\firstpage{1}
\htitle{Definitions of the Lambda Boo stars}
\hauthor{R. Faraggiana and M. Gerbaldi}
\title{Definitions of the Lambda Boo stars}
\author{R. Faraggiana \inst{1} \and M. Gerbaldi \inst{2,} \inst{3}}
\institute{Dipartimento di Astronomia, v. Tiepolo 11, I-34131 Trieste, Italy 
\and
Institut d'Astrophysique, 98bis, Bd. Arago, F-75014 Paris, France
\and Universit\'e de Paris-Sud XI}
\date{\today}
\maketitle
\begin{abstract}
Since the identification of these stars by Morgan et al. in 1943, various
definitions have been proposed for the stars of the Lambda Bootis group.
We present here the various definitions which have been given to these
objects in order to induce a general discussion on this topic.
\end{abstract}

\section{Introduction}
\label{intr}
The criteria to detect this class of peculiar A-type stars rely upon the choices
made by various authors in the last 50 years. Therefore several definitions of 
lambda Boo stars are found in the literature.

Both photometric and spectroscopic criteria have been used.
Some definitions so far proposed, concern only stars of spectral type near A0 
while others include A and F stars; no restrictions appear about the luminosity
class and therefore the evolutionary stage of the lambda Boo stars.

The common character of these stars, according to the various definitions, is 
the weakness of the metallic lines; however requirements such as high $v\sin i$
and deficiency of specific elements are introduced by some, but not all 
authors; the same remark concerns their kinematic properties.

We shall present below the criteria used by various authors in order to
understand the differences between the different lists of such stars published
up to now, and to open a discussion for the future.

\section{The discovery of the Lambda Boo Stars}

{\bf ... a  definition based on Spectral Classification  criteria ...}

Morgan et al. (1943) gave in fact implicitly the first definition of this group 
describing the peculiarities of Lambda Boo itself.

"The spectral type of Lambda Boo is near A0, as far as can it be determined. The 
spectral lines, while not unusually broad, are very weak, so that the only
 features easily visible are a weak K line and the Balmer series of hydrogen". 
However these authors did not define a group of stars; this was done later on  
from spectroscopy by Burbidge and Burbidge (1956).
Occasionally stars similar to Lambda Boo were discovered later by Slettebak. 
(Slettebak 1952 and 1954). 
 
{\bf ... and the first abundance analysis ...} 

Burbidge and Burbidge (1956) have analyzed Lambda Boo and 29 Cyg and they found
a metal deficiency for the elements Mg, Ca, Fe, Sc, Ti and Sr. Later on, Baschek
and Searle suggested that the oxygen abundance should be normal 
(quoted in the
Annual Report of the Director, Mount Wilson and Palomar Observatories, 1962-63
page 12). This was shown by Kodaira (1967) with infra-red spectra.

Very soon the photometry was used to detect Lambda Boo stars (Parenago 1958).

In 1965 Sargent (1965) showed that these stars can be distinguished from 
other weak lined stars such as horizontal branch stars by the fact that their
 space velocities are those of Population I stars and that they have moderately 
large rotational velocities.

In 1968, Settebak et al. introduced the first spectroscopic  definition of the
lambda Boo class : "These objects are defined spectroscopically as A-type stars
(as classified from Ca\,{\sc ii} K line to Balmer line ratio) with weakened 
metallic lines. They may be distinguished from other stars with the same 
characteristics (such as horizontal branch stars) by the fact that all show 
moderately large rotational velocities and small space velocities"

From an abundance analysis of 5 so-defined Lambda Boo stars,
Baschek and Searle (1969) found that only 3 of them (Lambda Boo, 29 Cygn and
$\pi^1$ Ori) form a group from the composition point of view; these authors 
suggested that these stars constitute a type of peculiar A stars. 
 
We recall that at that time, a list of 7 lambda Boo stars was available (see for
example Sargent, 1965), detected by spectroscopy or photometry. Occasionally a 
star with low abundances was discovered and often related to the lambda Bootis 
stars (see for example ADS 3910B in Sargent, 1966).

So from the beginning some confusion exists in the literature about which
objects should be considered as ``lambda Boo stars''. It is clear that
insufficient credit has been given to the true definition of this class found in
Baschek \& Searle (1969); ``...we define the Lambda Boo stars as stars whose 
composition resembles that of Lambda Boo itself...''.

\section{The eighties}
The lambda Boo stars were forgotten until the 1980s when both photometric and
spectroscopic researches underwent a revival of interest starting with the
paper by Hauck and Slettebak (1983) where the spectroscopic definition was 
expanded to the A-F stars.
Since then, different candidates have been selected by different criteria,
either photometric (for example: Hauck (1986) or spectroscopic (for example:
Abt 1984a).

At that time it became clear that the peculiarities of the Lambda Boo spectrum 
in the UV were easily detectable (Cucchiaro et al. 1980 ) even at the low TD1
resolution. Baschek et al. (1984) pointed out  that characteristic features
of the lambda Boo stars can be easily seen on low resolution IUE spectra.
Faraggiana et al. (1990) extended this research and defined the UV criteria
 useful to detect the lambda Boo stars in the range 120-200 nm. 

\section{An extensive spectroscopic classification}
An extensive spectroscopic classification has been made by Gray (1988) in the 
classical wavelength range.
Gray (1991) described in details the spectroscopic peculiarities detectable in 
the photographic domain at moderate resolution, providing a precise working
definition of the lambda Boo stars :
``Spectra of these stars are characterized by a weak Mg\,{\sc ii} 4481 line,
a K line type of A0 or slightly later and hydrogen-line type between A0 and F0.
For their hydrogen line type, the metallic-line spectrum is weak''. Moreover
their space velocities are those of Population I stars, and their rotational 
velocities are moderately high. The shape of the hydrogen lines profiles,
peculiar in some lambda Boo stars, is introduced as a further criterium to
separate two classes of these objects.

In 1990, Renson et al. collected all the stars that in the literature have been
called lambda Boo or candidate lambda Boo at least once, as well as objects 
called "weak-lines" stars that may be lambda Boo candidates. This catalogue
contains 101 stars.

The confusing situation is well illustrated by the two lists of lambda Boo
candidates extracted from the same sample of stars (the Bright Star Catalogue)
and based on similar classification by Abt (1984a), Abt and Morrell (1995) 
and by Gray (1988). Few stars are in common between these authors and some stars
classified as Lambda Boo by Abt are considered normal by
Gray and Garrison (1987).

On the basis of the selection by Gray (1988 and 1991) and UV criteria by
Faraggiana et al. (1990), a list of stars fulfilling the visible and/or UV
properties of lambda Boo itself has been established by
Gerbaldi and Faraggiana (1993). We consider all these stars to be reliable 
Lambda Boo candidates on the basis of the fact that
all the stars classified as Lambda Boo by Gray and observed by IUE belong to 
the same class according to UV criteria, and vice versa not one star among those
rejected on the basis of UV criteria appears as lambda Boo in Gray's list.

\section{The nineties}
The abundance determination by Venn and Lambert (1990), their interpretation
of the abundance pattern of the lambda Boo stars and the IR excess detected by 
IRAS around some lambda Boo stars (Sadakane and Nishida 1986) were the starting
points for observations in new directions : 

   - detection of lambda Boo stars in young clusters by Gray and Corbally (1993)
and by Levato et al.(1994).

   - detection of gas or dust shells around the Lambda Boo stars (Holweger \& 
Rentzsch-Holm 1995; Grady et al. 1996; King \& Patten 1992; Holweger 
\& St\"urenburg 1991; Bohlender \& Walker 1994; King 1994; Hauck et al. 1995;
Hauck et al. 1997)

   - new abundance analysis (St\"urenburg 1993) 

   - discussion on the position in the HR diagramme (Gerbaldi et al. 1993;
Iliev \& Barzova 1995)

   - observations of Lambda Boo candidates with asteroseismic techniques
(Paunzen et al. 1997) and detection of pulsating objects among them.

Charbonneau (1993) and Turcotte and Charbonneau (1993) computed diffusion
effects in the atmosphere of a lambda Boo stars giving time scale for the
duration of this phenomenon in the context of accreted material on the surface.

A new hypothesis on the origin of the Lambda Boo stars in the framework of the 
evolution of a close binary system has been recently developed by Andrievsky 
(1997).

The revival of ``theoretical'' interest for these stars prompted search for new 
candidates and it lead Gray (1997) to re-iterate and expand upon what he 
feels to be the best optical spectroscopic definition for the class of lambda
 Boo stars.

At the same time a new definition of the lambda Boo stars is given by Pauzen et
 al. (1997): `` Pop I hydrogen burning metal poor (except of C, N, O 
and S) A-type stars''; a list of 45 stars is given in their catalogue.

Only 8 stars among the 26 proposed by Abt (1984a) are
present in this catalogue. We notice also that Abt and Morrell (1995)
classified 46 lambda Boo stars, but only 9 are in common with the catalogue
by Paunzen et al. and for 9 other stars the classification by Paunzen et al.
as lambda Boo stars is not shared by Abt and Morrell.

Moreover, in the framework of spectral analysis of binary stars some authors
have classifed the nearly unseen component as a lambda Boo star (Griffin
et al. 1992; Hack et al. 1997).

\bigskip
\noindent
{\bf What will be the group of Lambda Boo stars in the next millennium ?}


\end{document}